\RequirePackage[2020-02-02]{latexrelease}

\documentclass{rQUF2e}

\usepackage{hyperref}
\usepackage{etoolbox}
\usepackage{kvoptions}
\usepackage{bbm}
\usepackage{natbib}
\usepackage{epstopdf}
\usepackage{booktabs}

\theoremstyle{plain}

\theoremstyle{definition}

\theoremstyle{remark}

\usepackage[table,xcdraw]{xcolor}
\usepackage{float}
\usepackage{algorithm}
\usepackage{algpseudocode}
\usepackage{placeins}

\makeatletter
\DeclareRobustCommand\ref{
  \@ifstar\@refstar\T@ref
}
\DeclareRobustCommand\pageref{
  \@ifstar\@pagerefstar\T@pageref
}
\makeatother

\begin{document}

\title{(Non-Parametric) Bootstrap Robust Optimization \\ for Portfolios and Trading Strategies}

\author{
DANIEL OLIVEIRA$^{\dag}$,
GROVER GUZMAN$^{\dag}$, and
NICK FIROOZYE$^{\ast}$\\
\affil{$\dag$Institute of Mathmatics and Statistics, Universidade de São Paulo, São Paulo, Brazil \\
$\ast$Department of Computer Science, University College London, London, UK}}

\maketitle

\begin{abstract}
Robust optimization provides a principled framework for decision-making under uncertainty, with broad applications in finance, engineering, and operations research. In portfolio optimization, uncertainty in expected returns and covariances demands methods that mitigate estimation error, parameter instability, and model misspecification. Traditional approaches—including parametric, bootstrap-based, and Bayesian methods—enhance stability by relying on confidence intervals or probabilistic priors but often impose restrictive assumptions. This study introduces a non-parametric bootstrap framework for robust optimization in financial decision-making. By resampling empirical data, the framework constructs flexible, data-driven confidence intervals without assuming specific distributional forms, thus capturing uncertainty in statistical estimates, model parameters, and utility functions. Treating utility as a random variable enables percentile-based optimization, naturally suited for risk-sensitive and worst-case decision-making. The approach aligns with recent advances in robust optimization, reinforcement learning, and risk-aware control, offering a unified perspective on robustness and generalization. Empirically, the framework mitigates overfitting and selection bias in trading strategy optimization and improves generalization in portfolio allocation. Results across portfolio and time-series momentum experiments demonstrate that the proposed method delivers smoother, more stable out-of-sample performance, offering a practical, distribution-free alternative to traditional robust optimization methods.
\end{abstract}


\newpage

\section{Introduction}

Robust optimization is a modern framework for decision making under uncertainty that offers an alternative to stochastic programming. Unlike stochastic methods, which rely on specific probabilistic assumptions, robust optimization makes more general assumptions about uncertainty, resulting in formulations that are both tractable and flexible \citep{fabozzi-etal-2012}. Originating in robust control engineering, the framework has since found wide application across finance, engineering, and operations research.

In portfolio optimization, decision variables typically represent portfolio weights, while uncertainty arises from key inputs such as expected returns and covariance matrices. Robust optimization addresses these challenges by constructing robust counterparts of the original problem, formulated as worst case scenarios over appropriately defined uncertainty sets. These sets are designed to capture deviations of parameters from their nominal values in a way that avoids overly conservative solutions. In doing so, robust optimization mitigates estimation errors, parameter uncertainty, and model misspecification, thereby promoting more stable and reliable decision making.

Existing approaches to robust portfolio optimization can be broadly grouped into three categories: classical robust optimization, bootstrap based robust optimization, and Bayesian robust optimization. Classical robust optimization follows a frequentist perspective, incorporating confidence intervals into the decision process to account for parameter uncertainty \citep{goldfarb-iyengar-2003, tutuncu2004robust, ceria-stubbs-2016}. Bootstrap based robust optimization extends this idea by constructing uncertainty sets from resampled confidence intervals rather than asymptotic estimates \citep{suspandi-etal-2017}. Bayesian robust optimization, in contrast, integrates prior beliefs and model uncertainty through posterior distributions, thereby framing robustness within a probabilistic updating scheme \citep{anderson2016, hansen2006}.

This study introduces a novel non-parametric bootstrap framework for robust optimization in financial decision making. Our methodology leverages bootstrap sample paths to construct flexible, data driven confidence intervals, avoiding the need for strong parametric assumptions about uncertainty sets. The framework can accommodate uncertainty in statistical estimates, model parameters, and utility functions. By treating utility as a random variable, it supports percentile based optimization, which provides a natural way to model uncertainty probabilistically while still addressing worst case scenarios. We demonstrate the effectiveness of this framework in portfolio optimization, following methodologies such as \cite{goldfarb-iyengar-2003, tutuncu2004robust, ceria-stubbs-2016}, and extend the analysis to parameter selection for trading strategies, building on the insights of \cite{harvey-liu-2014}.

Our approach connects to recent advances in robust optimization, reinforcement learning, and risk aware decision making, where percentile based criteria and multiple priors are increasingly used to control risk \citep{delage2010, cousins2023}. It also relates to the literature on overfitting in trading strategy design, where multiple testing and selection bias often lead to inflated backtest performance \citep{harvey-liu-2014, bailey-etal-2016}. By constructing bootstrapped confidence intervals around utility functions conditional on parameter choices, our framework provides a robust mechanism for addressing overfitting, ensuring that optimized strategies generalize more effectively to unseen data.

Against this backdrop, our study seeks to address three key questions. First, how does our proposed non parametric bootstrap based robust framework perform relative to classical and Bayesian alternatives in the portfolio optimization setting, where most robust methods have traditionally been applied? Second, can such a framework be generalized to other optimization tasks beyond portfolio allocation? To explore this, we apply our method to the novel problem of tuning trading strategy parameters. Third, how does our approach compare with standard train/test splits or empirical risk minimization techniques common in the machine learning literature?

The remainder of the paper is organized as follows. Section~2 reviews the relevant literature. Section~3 discusses robust optimization in the mean–variance setting, introducing robustness from a probabilistic perspective. Section~4 presents our proposed non-parametric bootstrap framework, with applications to portfolio construction and trading strategy optimization. Section~5 outlines the experimental setup, Section~6 reports empirical results, and Section~7 concludes with key insights and directions for future research.

\section{Related Work}

Robust control and portfolio optimization have been the focus of extensive research. Classical robust control methods were first introduced to portfolio optimization by \cite{tutuncu2004robust}, as well as \cite{ceria-stubbs-2016}. These works treated the mean-variance optimization (MVO) problem as a maximin problem, incorporating confidence intervals for parameter estimates. This frequentist approach showed robust performance in various applications and has been widely cited in the literature \citep{kim2018robust}.

To address challenges with covariance matrix estimation, regularization and shrinkage methods such as those proposed by \cite{ledoit-2004} have been incorporated into robust portfolio frameworks. \cite{lopezdeprado-2016} and \cite{pedersen-2021} further explored practical implementations of robust methods that mitigate overfitting in asset allocation.

Bootstrap methods have gained traction as an alternative to frequentist approaches. \cite{suspandi-etal-2017} applied parametric bootstrap estimates to derive confidence intervals for MVO parameters, showcasing the potential of bootstrap techniques in enhancing decision-making under uncertainty. \cite{gravell2020} extended bootstrap methodologies to adaptive robust control frameworks, employing bootstrap confidence intervals in linear-quadratic regulators.

For covariance clustering and robust optimization, \cite{bongiorno-2022} introduced hierarchical clustering with bootstrap confidence intervals. Their work demonstrated the efficacy of clustering-based MVO models in improving portfolio performance by reducing estimation noise. Similar clustering-based methods have been applied in conjunction with regularized covariance estimates for enhanced robustness \citep{lopezdeprado-2016}.

Bayesian robust control provides another perspective by incorporating parameter and model uncertainty through hierarchical Bayesian frameworks. \cite{anderson2016} proposed a robust Bayesian portfolio choice model, integrating multiple priors to account for uncertainty in model assumptions. \cite{delage2010} introduced a percentile criterion approach for handling uncertainty in Markov Decision Processes (MDPs), which has since been extended to reinforcement learning by \cite{cousins2023}. These methods align closely with Bayesian control under uncertainty, as initially discussed by \cite{hansen2006} and formalized in maximin utility models by \cite{gilboa1989}.

Dependent bootstrap techniques for time-series data have also been explored extensively. \cite{lahiri2003} provided a comprehensive review of block bootstrap methods, including stationary and circular block bootstrap, which are commonly used for dependent financial data. \cite{kreiss2014} extended these ideas to locally stationary processes, highlighting theoretical guarantees of bootstrap-based inference for time-series analysis. \cite{dahl2022} recently introduced a novel GAN-based bootstrap framework, providing an alternative to traditional block bootstrap approaches.

Finally, statistical testing methods such as the Reality Check test by \cite{white-2000} and the Model Confidence Set by \cite{hansen-2011} offer robust frameworks for hypothesis testing in financial strategy evaluation. These tests complement bootstrap methods by ensuring statistical significance in performance evaluation \citep{kreiss2011, kreiss2023}.

\section{Setting the Stage}

Robust optimization is a framework for decision-making under uncertainty, particularly useful in portfolio optimization, where model misspecification and estimation errors can significantly impact performance. A key aspect of robust optimization is the explicit treatment of uncertainty in key parameters, such as the expected return vector $\boldsymbol{\mu}$ and the covariance matrix $\mathbf{\Sigma}$, ensuring that decisions remain resilient to adverse scenarios.

This section introduces the fundamental concepts of robust portfolio optimization and explores its various formulations. We begin by defining the core problem of mean-variance optimization (MVO) under uncertainty and present different approaches to handling parameter uncertainty, including classical robust optimization, Bayesian robust optimization, and bootstrap-based robust optimization. We then unify the notation across these methodologies and establish connections between different robustness paradigms.

\subsection{Mean-Variance Optimization under Uncertainty}

Classical robust optimization provides a frequentist framework for handling parameter uncertainty in decision-making problems. In the standard mean-variance optimization (MVO) setting, the objective is to identify the portfolio weight vector $\boldsymbol{w} \in \mathbb{R}^{d \times 1}$ that maximizes the expected utility of an investor.

\begin{equation}
\boldsymbol{w}^{*} = \arg\max_{\boldsymbol{w}} \Bigg[ \boldsymbol{\mu}^\top \boldsymbol{w} - \frac{\lambda}{2} \boldsymbol{w}^\top \mathbf{\Sigma} \boldsymbol{w} \Bigg],
\end{equation}

where $\lambda > 0$ denotes the investor’s risk-aversion parameter, $\boldsymbol{\mu} \in \mathbb{R}^{d \times 1}$ the expected returns, and $\mathbf{\Sigma} \in \mathbb{R}^{d \times d}$ is the covariance matrix of the asset returns.

In practice, the true values of $\boldsymbol{\mu}$ and $\mathbf{\Sigma}$ are unknown and must be estimated from historical data, leading to estimation errors. A key issue with the MVO framework is its sensitivity to such estimation errors. Small perturbations in the inputs can result in large shifts in the optimal portfolio weights, rendering the solutions unstable. This problem is well documented in the literature: \citet{chopra-1993} and \citet{michaud-1998} show that minor changes in input parameters can dramatically affect portfolio allocations. \citet{best-grauer-1991} specifically highlights the sensitivity of MVO solutions to changes in expected return estimates. Furthermore, empirical findings by \citet{jobson-korkie-1981} demonstrate that naive equal-weight portfolios can outperform optimized MVO portfolios in terms of Sharpe ratios. Estimation errors are not limited to expected returns; as emphasized by \citet{bailey-lopezdeprado-2012} and \citet{lopezdeprado-2016}, the sample covariance matrix is often numerically ill-conditioned, meaning that small errors can lead to drastic changes in its eigenstructure, severely distorting portfolio construction.

To mitigate these issues, classical robust optimization constrains the estimation of ${\boldsymbol{\mu}}$ and/or ${\mathbf{\Sigma}}$ within an uncertainty set $\mathbf{U}({\boldsymbol{\mu}}, {\mathbf{\Sigma}})$, leading to a maximin formulation:

\begin{equation}
\boldsymbol{w}^{*} = \arg\max_{\boldsymbol{w}} \min_{\boldsymbol{\mu},\mathbf{\Sigma} \in \mathbf{U}({\boldsymbol{\mu}},{\mathbf{\Sigma}})} \Bigg[ \boldsymbol{\mu}^\top \boldsymbol{w} - \frac{\lambda}{2} \boldsymbol{w}^\top \mathbf{\Sigma} \boldsymbol{w} \Bigg].
\end{equation}

A common specification of the uncertainty set for $\boldsymbol{\mu}$ is a quadratic region:

\begin{equation}
\mathbf{U}(\boldsymbol{\mu}) = \left\{ \boldsymbol{\mu} \; \Big| \; (\boldsymbol{\mu} - \hat{\boldsymbol{\mu}})^\top \Omega_{\boldsymbol{\mu}}^{-1} (\boldsymbol{\mu} - \hat{\boldsymbol{\mu}}) \leq \kappa^2 \right\},
\end{equation}

where $\hat{\boldsymbol{\mu}} \in \mathbb{R}^{d \times 1}$ is the estimated mean return vector, $\Omega_{\boldsymbol{\mu}} \in \mathbb{R}^{d \times d}$ represents the covariance matrix of the estimates, and $\kappa^2$ controls the size of the uncertainty set.

This formulation closely aligns with confidence regions under a multivariate normal assumption. Suppose $\boldsymbol{\mu} \sim \text{MVN}(\hat{\boldsymbol{\mu}}, \Omega_{\boldsymbol{\mu}})$, where $\hat{\boldsymbol{\mu}}$ denotes the estimator of the mean and $\Omega_{\boldsymbol{\mu}}$ is the associated estimation uncertainty. Then the quadratic form

\begin{equation}
Y = (\boldsymbol{\mu} - \hat{\boldsymbol{\mu}})^\top \Omega_{\boldsymbol{\mu}}^{-1} (\boldsymbol{\mu} - \hat{\boldsymbol{\mu}}) \sim \chi^2_d
\end{equation}

follows a Chi-squared distribution with $d$ degrees of freedom. The uncertainty level $\kappa^2$ can thus be calibrated to a confidence level $\gamma = 1 - \alpha$ as:

\begin{equation}
\mathbb{P}(Y \leq \kappa^2) = F_Y(\kappa^2) = 1 - \alpha \quad \Rightarrow \quad \kappa^2 = F_Y^{-1}(1 - \alpha),
\end{equation}

where $F_Y(\cdot)$ is the cumulative distribution function (CDF) of the $\chi^2_d$ distribution. This relationship allows us to interpret the uncertainty set $\mathbf{U}(\boldsymbol{\mu})$ as a confidence region with level $\gamma$, offering a statistical justification for its use in robust optimization.

This formulation, proposed by \cite{tutuncu2004robust} and \cite{ceria-stubbs-2016}, ensures that the portfolio is optimized for the worst-case parameter estimates within the given confidence intervals. Empirical results confirm that this approach enhances stability and robustness \citep{kim2018robust}.

An alternative to analytically derived confidence regions is to employ bootstrap methods to estimate uncertainty sets in a fully non-parametric manner. Unlike traditional approaches that rely on asymptotic approximations or distributional assumptions (e.g., normality), bootstrap-based robust optimization constructs empirical confidence regions directly from resampled data. This method enables modeling uncertainty based on the observed data distribution, capturing higher-order moments, dependencies, and finite-sample effects.

Following a similar approach to the one proposed by \citet{suspandi-etal-2017}, the robust mean-variance optimization problem can be reformulated as:

\begin{equation}
\boldsymbol{w}^* = \arg\max_{\boldsymbol{w}} \min_{\boldsymbol{\mu},\mathbf{\Sigma} \in CI^*_\gamma(\boldsymbol{\mu},\mathbf{\Sigma})} \Bigg[ \boldsymbol{\mu}^\top \boldsymbol{w} - \frac{\lambda}{2} \boldsymbol{w}^\top \mathbf{\Sigma} \boldsymbol{w} \Bigg],
\end{equation}

where $CI^*_\gamma(\boldsymbol{\mu}, \mathbf{\Sigma})$ denotes empirical confidence regions obtained from bootstrap replications at level $\gamma$:
\begin{equation}
CI^*_\gamma(\boldsymbol{\mu}, \mathbf{\Sigma}) = CI_\gamma\left( \{\mathbb{E}[X^*]\}, \{\mathbb{V}[X^*]\} \right).
\end{equation}
Here, $X^*$ denotes a bootstrap sample drawn using a method such as the block or stationary bootstrap, which is designed to preserve the serial dependence inherent in financial time series \cite{lahiri2003}. For each resampled dataset, we compute empirical estimates of the mean and covariance matrix. These estimates are then used to construct the joint bootstrap confidence region $CI^*_\gamma(\boldsymbol{\mu}, \mathbf{\Sigma})$ at the desired confidence level $\gamma$. The following pseudo-code outlines the procedure:

\begin{algorithm}[H]
\caption{Bootstrap-Based Robust Mean-Variance Optimization}
\label{alg:algo1}
\begin{algorithmic}[1]

\State $\gamma \gets$ confidence level (e.g., 0.75 or 0.90)
\State $\{X_t\} \gets$ $(T \times d)$-dimensional time series data
\For{$i = 1$ to $S$} \Comment{Generate $S$ bootstrap replications}
    \State $X^*_i \gets \text{BootstrapSample}(X)$
    \State $\boldsymbol{\mu}^*_i \gets \mathbb{E}[X^*_i]$
    \State $\mathbf{\Sigma}^*_i \gets \mathbb{V}[X^*_i]$
\EndFor

\State $CI^*_\gamma(\boldsymbol{\mu}^*, \mathbf{\Sigma}^*) = CI_\gamma\left( \{\boldsymbol{\mu}^*_i\}_{i=1}^S, \{\mathbf{\Sigma}^*_i\}_{i=1}^S \right)$
\Comment{Construct empirical confidence regions}

\State $\boldsymbol{\theta}^* \gets \arg\max_{\boldsymbol{\theta}} \Bigg(
\min_{\boldsymbol{\mu}, \mathbf{\Sigma} \in CI^*_\gamma(\boldsymbol{\mu}^*, \mathbf{\Sigma}^*)} 
\Big[ \boldsymbol{\theta}^\top \boldsymbol{\mu} - \frac{\lambda}{2} \boldsymbol{\theta}^\top \mathbf{\Sigma} \boldsymbol{\theta} \Big] \Bigg)$

\end{algorithmic}
\end{algorithm}

\citet{gravell2020} extends robust optimization to the adaptive control setting by incorporating bootstrap-based confidence intervals into a robust Linear Quadratic Regulator (LQR) framework. This approach enables the control strategy to adapt over time as new data arrives, using resampling techniques to continually update uncertainty estimates.

Bayesian robust optimization, in contrast, adopts a probabilistic perspective by modeling parameter uncertainty via prior distributions. Letting $\boldsymbol{\theta} = (\boldsymbol{\mu}, \mathbf{\Sigma})$ denote the model parameters with prior distribution $p(\boldsymbol{\theta})$, Bayes’ theorem yields the posterior distribution conditioned on observed data $D$:

\begin{equation}
p(\boldsymbol{\theta} \mid D) \propto p(D \mid \boldsymbol{\theta}) \, p(\boldsymbol{\theta}).
\end{equation}

This leads to the predictive distribution of asset returns:

\begin{equation}
p(R \mid D) = \int p(R \mid \boldsymbol{\theta}) \, p(\boldsymbol{\theta} \mid D) \, d\boldsymbol{\theta}.
\end{equation}

A standard Bayesian portfolio optimization problem then seeks to maximize expected utility under this predictive distribution:

\begin{equation}
\max_{\boldsymbol{w}} \; \mathbb{E}[U(R; \boldsymbol{w})] = \max_{\boldsymbol{w}} \int U(R; \boldsymbol{w}) \, p(R \mid D) \, dR.
\end{equation}

However, expectation-based objectives can be vulnerable to tail risk. A more robust alternative is to optimize a utility-based risk measure, such as a lower percentile of the utility distribution:

\begin{equation}
L(R; \boldsymbol{w}) = P_\alpha[U(R, \boldsymbol{w})],
\end{equation}

where $P_\alpha(\cdot)$ denotes the $\alpha$-th percentile. \citet{delage2010} explore this idea through chance constraints, which ensure that specified outcomes occur with high probability. These techniques have also been extended to reinforcement learning, where risk-sensitive decision-making under uncertainty is critical \citep{cousins2023}.

A particularly expressive Bayesian formulation introduces uncertainty over the prior itself, resulting in a hierarchical Bayesian model. Instead of assuming a fixed prior, we define a family of priors governed by hyperparameter vector $\boldsymbol{\eta}$:

\begin{equation}
p(R \mid D, \boldsymbol{\eta}) = \int p(R \mid \boldsymbol{\theta}) \, p(\boldsymbol{\theta} \mid D, \boldsymbol{\eta}) \, d\boldsymbol{\theta}.
\end{equation}

The robust decision-making problem can then be framed as a max-min optimization over the hyperparameter space:

\begin{equation}
\max_{\boldsymbol{w}} \; \min_{\boldsymbol{\eta}} \; \mathbb{E}^{\boldsymbol{\eta}}[U(R; \boldsymbol{w})] = \max_{\boldsymbol{w}} \; \min_{\boldsymbol{\eta}} \int U(R; \boldsymbol{w}) \, p(R \mid D, \boldsymbol{\eta}) \, dR.
\end{equation}

This robust Bayesian formulation has been explored through various lenses, including adversarial learning, model averaging, and ambiguity-averse decision-making \citep{gilboa1989, hansen2006, anderson2016}.

In summary, classical robust optimization emphasizes worst-case guarantees via confidence regions and deterministic constraints. Bootstrap-based methods offer a flexible, non-parametric alternative rooted in empirical distributions. Bayesian robust optimization integrates prior beliefs and models uncertainty probabilistically. Each framework offers distinct advantages—be it interpretability, adaptability, or probabilistic expressiveness—and the appropriate choice depends on the specific context and nature of uncertainty in the decision-making problem.

\section{Proposed Methodology}

We propose a robust optimization framework that addresses both parameter and model uncertainty using a non-parametric bootstrap approach. Unlike traditional methods that rely on parametric assumptions or asymptotic theory, our framework constructs empirical distributions directly from the data, making it model-agnostic and highly adaptable. A key innovation is to treat utility as a random variable with an associated distribution, enabling optimization over distributional functionals—such as its $\alpha$-th percentile.

The methodology begins by generating bootstrap samples (e.g., block, circular, or stationary bootstrap) from the original time series data, thereby preserving temporal dependencies common in financial settings. For each bootstrap replication, we compute the corresponding utility. The optimization objective is then defined based on a selected percentile of the utility distribution—e.g., the 25th percentile—which provides a principled mechanism to manage downside or worst-case risk.

Formally, let $\alpha = 0.25$, and define $P_\alpha(X)$ as the $\alpha$-th percentile of a random variable $X$. The robust optimization problem can be formulated as:

\begin{equation}
\max_{\boldsymbol{\theta}} \; P_\alpha \Big( U(X^* ; \boldsymbol{\theta}) \Big) = \max_{\boldsymbol{\theta}} \; P_\alpha \Big( \boldsymbol{\theta}^\top \boldsymbol{\mu}^* - \frac{\lambda}{2} \boldsymbol{\theta}^\top \mathbf{\Sigma}^* \boldsymbol{\theta} \Big),
\end{equation}

where $\boldsymbol{\theta}$ denotes the decision variables (e.g., $\boldsymbol{\theta} := \boldsymbol{w}$ in the mean-variance setting), and $\boldsymbol{\mu}^*$ and $\mathbf{\Sigma}^*$ are the bootstrapped estimates of the mean vector and covariance matrix, respectively, computed as described in Algorithm~\ref{alg:algo1}.

This formulation enables explicit control over the lower tail of the utility distribution, yielding solutions that are robust to adverse outcomes potentially hidden by expectation-based objectives.

Although this approach shares some similarities with parametric bootstrap confidence intervals (CIs), the key distinction lies in the treatment of dependence. Parametric bootstrap methods typically assume independence between bootstrapped quantities, whereas our framework preserves potential dependencies between realizations of $\boldsymbol{\mu}^*$ and $\mathbf{\Sigma}^*$. This added flexibility broadens the scope of the analysis and more accurately reflects the joint distributional uncertainty. Here, $\boldsymbol{\theta}$—representing portfolio weights—can be interpreted as a hyperparameter of the underlying decision strategy.

To generalize, consider a trading strategy defined by a sequence of positions $\{\boldsymbol{w}_t(\boldsymbol{\theta})\}$, where $\boldsymbol{w}_t$ depends on the information set at time $t^-$ and a vector of decision variables, now relevant hyperparameters for the problem at hand, $\boldsymbol{\theta}$. For instance, in the context of time-series momentum strategies, $\boldsymbol{\theta}$ may include the lookback window used to calculate cumulative past returns. The associated profit and loss (P\&L) process, assuming initial wealth $V_0$ and linear transaction costs, is given by:

\begin{equation}
V_t = V_0 + \sum \boldsymbol{w}_t(\boldsymbol{\theta})^\top \Delta \boldsymbol{p}_t - \sum \text{tc} \times \big| \Delta \boldsymbol{w}_t(\boldsymbol{\theta}) \big|,
\end{equation}

where $\Delta \boldsymbol{p}_t \in \mathbb{R}^{d \times 1}$ denotes the vector of asset price changes, and $\text{tc}$ represents a vector of transaction costs.

The goal is to select hyperparameters $\boldsymbol{\theta}$ that maximize a performance functional, such as the Sharpe Ratio:

\begin{equation}
\max_{\boldsymbol{\theta}} \; \text{SR}\Big( \Delta V(\boldsymbol{\theta}) \Big),
\end{equation}

though this empirical optimization is susceptible to overfitting on historical data. This issue has been extensively studied in the machine learning and statistical literature in the context of empirical risk minimization (ERM), where models are trained to optimize performance on observed data \citep{bagnell-2005, bental-etal-2013, duchi-etal-2016, sagawa-etal-2020}.

To mitigate overfitting and overreliance on potentially spurious correlations that may not generalize, robust optimization techniques have been developed \cite{bagnell-2005}. A key idea in this literature is to replace the empirical objective with a distributionally robust counterpart that controls for the worst-case expected performance over an uncertainty set $\mathcal{Q}$ of plausible data-generating distributions. This leads to the following formulation:

\begin{equation}
\min_{\boldsymbol{\theta} \in \Theta} \; \sup_{Q \in \mathcal{Q}} \; \mathbb{E}_{(x,y) \sim Q} \left[ \ell(\boldsymbol{\theta}; (x,y)) \right],
\end{equation}

where $\mathcal{Q}$ encodes the family of distributions that capture allowable perturbations or shifts from the empirical distribution. Choosing $\mathcal{Q}$ carefully — e.g., as a divergence ball around the empirical distribution — can enhance robustness to distributional shifts and adversarial conditions, but may also lead to overly conservative models if $\mathcal{Q}$ is too broad \citep{duchi-etal-2016}.

In robust learning, performance is often evaluated via worst-group criteria, where data are partitioned into subgroups (e.g., environments, regimes, or clusters) and robustness is assessed by minimizing the maximum risk across these groups. This ensures that the selected hyperparameters perform reliably even in challenging or underrepresented market scenarios. Such percentile- or group-based objectives are well aligned with the methodology proposed in this paper, which seeks to optimize functionals of the bootstrapped utility distribution—rather than its mean.

To further mitigate overfitting, we propose applying the block bootstrap to the return vector process $\boldsymbol{p}_t$, denoted $\boldsymbol{p}^*_t$. This allows us to define bootstrapped strategy weights as $\boldsymbol{w}_t := \boldsymbol{w}_t(\boldsymbol{p}^*_{t}; \boldsymbol{\theta})$ for a fixed hyperparameter vector $\boldsymbol{\theta}$, yielding a bootstrapped P\&L process $\Delta V_t^*(\boldsymbol{\theta})$. The distribution of strategy performance under this bootstrapped setup enables the construction of robustified metrics, such as:

\begin{equation}
\text{SR}^*(\boldsymbol{\theta}) = \text{SR}\Big(\Delta V_t^*(\boldsymbol{\theta})\Big),
\end{equation}

and ultimately allows for optimizing robust percentiles or other distributional functionals over $\boldsymbol{\theta}$.

This setup enables automated, robust optimization of trading strategies with predefined hyperparameters. Examples include decision rules based on lookback windows, threshold parameters, or return scaling using single-period mean-variance optimization with transaction costs. For instance, the bootstrapped utility function for a single asset can be expressed as:
$$
U^*(\theta) = \Big\{ \theta{w}\mu^* - \frac{\lambda}{2} \sigma^{*2} w^2 - \text{tc}^* \times |\Delta{w}| \Big\},
$$
where $\theta$ scales the return forecast, $\lambda$ denotes risk aversion, and $\mu^*$ and $\sigma^*$ are bootstrapped estimates of mean return and volatility, respectively—each based on dependent bootstrap samples $p_t^*$. Transaction costs $\text{tc}^*$ and return estimates are also bootstrapped. As noted in \citet{grinold1999}, forecast scaling is often used in practice to avoid over- or under-trading.

By leveraging the entire bootstrap ensemble, a density for $U^*(\theta)$ is generated, allowing explicit manipulation of the objective function based on the hyperparameter choices $\theta$ and the desired functional to control.

For instance, an objective function might focus on a specific percentile. Rather than optimizing for the mean Sharpe Ratio, this framework enables optimizing for the 25th percentile, ensuring robust performance across 75\% of the bootstrapped scenarios. This percentile-based approach prioritizes resilience, delivering strategies that perform well under most bootstrapped outcomes, rather than excelling only under average conditions.

In Algorithm~\ref{alg:algo2}, we present the general framework underlying our proposed methodology. The algorithm implements a non-parametric, dependence-preserving bootstrap procedure for robust optimization. It applies to a wide range of estimation and control tasks where parameter uncertainty plays a central role. The process begins by generating a $S$ bootstrap samples $\{X^*_i\}_{i=1}^S$ from the original time series ${X_t}$. For each candidate hyperparameter vector $\boldsymbol{\theta} \in \Xi$, we compute the utility $U_i(\boldsymbol{\theta})$ over each resampled dataset $X_i^*$, producing an empirical distribution of utility values conditional on $\boldsymbol{\theta}$.

From this distribution, we extract a robust summary statistic or functional—such as the mean, the skewness, or the $\alpha$-th percentile—denoted by $V\Big(\{U_i^*(\boldsymbol{\theta})\}_{i=1}^S\Big)$. The final optimization step involves selecting the hyperparameter configuration $\boldsymbol{\theta}$ that maximizes this robust functional. This framework allows one to directly target distributional properties of interest, enabling robust decisions that are less sensitive to estimation error and adverse outcomes.

\begin{algorithm}[H]
\caption{An algorithm for Non-parametric Bootstrap Robust Optimization}
\label{alg:algo2}
\begin{algorithmic}[1] 

\State $\{X_t\} \gets$ $(T \times d)$-dimensional time series data
\State $\{X_i^*\} \gets$ $S$ bootstrap samples of the time-series data 
\While{$i \leq S$}
    \State $X^*_i \gets \hbox{bootstrap}(\{X_t\},i) $
    \State $i \gets i + 1$
\EndWhile

\State $\{\boldsymbol{\theta}\} \gets$ a list of ($k$-dim) vectors of relevant hyperparameters
\State $\{U^*(\boldsymbol{\theta})\} \gets$ a list of $N$ statistic of the bootstrap utilities associated with $\boldsymbol{\theta}$
\For{$\boldsymbol{\theta} \in \Xi$}
    \While{$j \leq N$}
        \State $\{U_i^*(\boldsymbol{\theta})\} \gets$ a list with $S$ utilities associated with the given hyperparameter vector $\boldsymbol{\theta}$ 
    \EndWhile
    \State $U^*(\boldsymbol{\theta}) \gets \mathbb{E}\Big[\{U_i^*(\boldsymbol{\theta})\}\;\Big]$
\EndFor
\State Find $\boldsymbol{\theta} \gets \arg\max_{\boldsymbol{\theta}} ~V\Big(\{U^*(\boldsymbol{\theta})\}\Big) := P_{\alpha}\Big(\{U^*(\boldsymbol{\theta})\}\Big)$

\end{algorithmic}
\end{algorithm}

Returning to the case of robust MVO, we now apply the general procedure to construct the utility distribution governed by the hyperparameter $\boldsymbol{\eta}_j := \boldsymbol{\theta}$. The relevant utility function for each bootstrap sample is given by:

\begin{equation}
U_i^*(\boldsymbol{\theta}) = U(X_i^*; \boldsymbol{\theta} ) = \boldsymbol{\theta}^{\top} \mathbb{E}[X_i^*] - \frac{\lambda}{2} \boldsymbol{\theta}^{\top} \mathbb{V}[X_i^*] \boldsymbol{\theta}.
\end{equation}

This yields the empirical utility distribution $\{U_i^*(\boldsymbol{\theta})\}_{i=1}^{S}$, conditional on a fixed portfolio weight vector $\boldsymbol{\theta}$. Accordingly, we define the objective functional as:

\begin{equation}
V\Big(\{U_i^*(\boldsymbol{\theta})\}_{i=1}^{S}\Big) = P_\alpha\left(\{U_i^*(\boldsymbol{\theta})\}\right),
\end{equation}

where $P_\alpha$ denotes the $\alpha$-th percentile of the utility distribution. The optimization objective then becomes to maximize this robust measure of utility. Alternatively, we may choose to maximize expected utility subject to a chance constraint on the minimum acceptable performance:

\begin{align}
\max_{\boldsymbol{\theta}} \quad & \mathbb{E}[U_i^*(\boldsymbol{\theta})] \\
\text{subject to} \quad & P_\alpha\left(\{U_i^*(\boldsymbol{\theta})\}\right) \geq c.
\end{align}

Specializing to this MVO setting, we derive the following algorithm:

\begin{algorithm}[H]
\caption{Non-parametric Bootstrap Robust MVO via Gradient Ascent}
\label{alg:algo3}
\begin{algorithmic}[1]

\State $\alpha \gets$ relevant percentile
\State $\nu \gets$ learning rate
\State $\{X_t\} \gets$ $(T \times d)-$dimensional time series data
\State $i \gets 1$
\While{$i \leq S$}
    \State $X^*_i \gets \text{bootstrap}(X, i)$
    \State $\Theta^*_i \gets (\mathbb{E}[X^*_i], \mathbb{V}[X^*_i])$
    \State \;\;\;\; $\equiv (\boldsymbol{\mu}_i^*, \mathbf{\Sigma}_i^*)$
    \State $i \gets i + 1$
\EndWhile

\While{stopping criterion not met}
    \State $i \gets 1$
    \While{$i \leq S$}
        \State $U_i^*(\boldsymbol{\theta}) \gets \boldsymbol{\theta}^\top \boldsymbol{\mu}_i^* - \frac{\lambda}{2} \boldsymbol{\theta}^\top \mathbf{\Sigma_i}^* \boldsymbol{\theta}$
        \State $i \gets i + 1$
    \EndWhile
    
    \State $p \gets P_\alpha\left(\{U_i^*(\boldsymbol{\theta})\}_{i=1}^S\right)$
    \State $\nabla_{\boldsymbol{\theta}} p \gets \frac{\partial}{\partial \boldsymbol{\theta}} P_\alpha\left(\{U_i^*(\boldsymbol{\theta})\}_{i=1}^S\right)$
    \State $\boldsymbol{\theta} \gets \boldsymbol{\theta} + \nu \cdot \nabla_{\boldsymbol{\theta}} p$
\EndWhile

\end{algorithmic}
\end{algorithm}

Thus, Algorithm \ref{alg:algo3} uses gradient ascent to solve the robust optimization problem:

\begin{align*}
\boldsymbol{\theta}^* &\gets \arg\max_{\boldsymbol{\theta}} \; P_\alpha\left(\left\{ \boldsymbol{\theta}^\top \boldsymbol{\mu}_i^* - \frac{\lambda}{2} \boldsymbol{\theta}^\top \mathbf{\Sigma}_i^* \boldsymbol{\theta} \right\}_{i=1}^{S} \right) \\
&= \arg\max_{\boldsymbol{\theta}} \; P_\alpha\left(\left\{ U_i^*(\boldsymbol{\theta}) \right\}_{i=1}^{S} \right).
\end{align*}

We now turn to the case of trading strategies parameterized by a vector of hyperparameters $\boldsymbol{\theta} \in \Xi$. These strategies define dynamic position rules $\boldsymbol{w}_t(\boldsymbol{\theta})$ that depend on information available at time $t$ and may include features such as lookback windows, thresholds, or signal transformations. In this context, we seek to evaluate the robustness of different configurations $\boldsymbol{\theta}$ by generating an empirical utility distribution through bootstrapping and optimizing a functional of that distribution, such as its $\alpha$-th percentile.

Following Algorithm \ref{alg:algo4}, we first construct a bootstrap samples $\{X_i^*\}_{i=1}^S$ from the original price data $\{X_t\}_{t=1}^{T}$. For each hyperparameter configuration $\boldsymbol{\theta} \in \boldsymbol{\Xi}$ and each bootstrap path $X_i^{*}$, we compute a trading signal $S_{t,i}$, a forecast $F_{t,i}$, and a backtested return stream $R_{t,i}$. These are used to evaluate utility $U_i^*(\boldsymbol{\theta})$—for instance, cumulative return, Sharpe ratio, or mean-variance utility—resulting in a utility distribution $\{U_i^*(\boldsymbol{\theta})\}_{i=1}^S$ for each $\boldsymbol{\theta}$. A robust summary functional, such as the empirical average or a lower percentile, is computed as $U^*(\boldsymbol{\theta}) := \mathbb{E}[{U_i^*(\boldsymbol{\theta})}]$, and the final optimization problem consists of maximizing the $\alpha$-th percentile of these summary utilities over all hyperparameter choices

$$
\boldsymbol{\theta}^* = \arg\max_{\boldsymbol{\theta} \in \Xi} ~P_\alpha\Big( { U^*(\boldsymbol{\theta}) } \Big).
$$

\begin{algorithm}[H]
\caption{Non-parametric Bootstrap Robust Strategy Optimization}
\label{alg:algo4}
\begin{algorithmic}
\State $\alpha \gets$ relevant percentile
\State $\{X_{t}\} \gets$ $(T \times d)$-dimensional time series data
\State $\Xi \gets$ set of hyperparameters that characterize a trading signal
\State $i \gets 1$
\While{$i \leq S$}
    \State $X^*_i \gets \text{bootstrap}(X, i)$
    \State $i \gets i + 1$
\EndWhile
\For{each $\boldsymbol{\theta} \in \Xi$}
    \State $i \gets 1$
    \While{$i \leq S$}
        \State $S_{t,i} \gets \text{signal}(X^{*}_i, \boldsymbol{\theta})$
        \State $F_{t,i} \gets \text{forecast}(S_{t,i})$
        \State $R_{t,i} \gets \text{backtest}(X^{*}_i, F_{t,i})$
        \State $\{U^*_{i}(\boldsymbol{\theta})\} \gets \text{utility}(R_{t,i})$
        \State $i \gets i + 1$
    \EndWhile
    \State $U^*(\boldsymbol{\theta}) \gets \mathbb{E}\Big[\{U^*_{i}(\boldsymbol{\theta})\}\Big]$
\EndFor
\State $\boldsymbol{\theta}^{*} \gets \arg\max_{\boldsymbol{\theta} \in \Xi} ~ P_{\alpha}\Big(\{U^*(\boldsymbol{\theta})\}\Big)$
\end{algorithmic}
\end{algorithm}

\section{Experiment Setup}

This section describes the empirical setup employed to evaluate the proposed methodology. We consider two tasks: (i) portfolio optimization within the classical mean–variance framework, and (ii) the construction of a time-series momentum strategy. For the portfolio optimization experiment, we use ETF data obtained from the WRDS database, while for the time-series momentum experiment we rely on both ETF and futures data, with the latter sourced from the Pinnacle database. We also outline how our proposed model is evaluated throughout the experiments, the benchmarks employed for comparison, and the metrics used to assess robustness and performance.

\subsection{Dataset}
\label{sec:benchmarks}

\subsubsection{Exchange-Traded Funds (ETFs)}

Our empirical analysis relies on a panel of Exchange-Traded Funds (ETFs) with historical price data obtained from the WRDS database. The dataset covers multiple asset classes, including equities, fixed income, commodities, and volatility, providing a broad basis for the evaluation of portfolio and trading strategies. The data are sampled at a daily frequency, beginning on April 10, 2006. In the panel structure, each row corresponds to a trading day and each column records the closing price of a specific ETF.

Table~\ref{tab:etf-descriptions} provides the list of ETF tickers included in the study along with a brief description of the market exposure or sector each one represents.

\begin{table}[H]
\centering
\caption{Description of ETFs used in the dataset}
\label{tab:etf-descriptions}
\begin{tabular}{ll}
\toprule
\textbf{Ticker} & \textbf{Description} \\
\midrule
SPY  & S\&P 500 Index ETF \\
IWM  & Russell 2Small-Cap ETF \\
EEM  & Emerging Markets Equity ETF \\
TLT  & Long-Term U.S. Treasury Bonds ETF \\
USO  & Crude Oil Futures ETF \\
GLD  & Gold Trust ETF \\
XLF  & Financial Sector ETF \\
XLB  & Materials Sector ETF \\
XLK  & Technology Sector ETF \\
XLV  & Healthcare Sector ETF \\
XLI  & Industrials Sector ETF \\
XLU  & Utilities Sector ETF \\
XLY  & Consumer Discretionary Sector ETF \\
XLP  & Consumer Staples Sector ETF \\
XLE  & Energy Sector ETF \\
VIX  & CBOE Volatility Index ETF \\
AGG  & Aggregate Bond Market ETF \\
DBC  & Commodities Index Tracking ETF \\
HYG  & High-Yield Corporate Bonds ETF \\
LQD  & Investment-Grade Corporate Bonds ETF \\
UUP  & U.S. Dollar Index ETF \\
\bottomrule
\end{tabular}
\end{table}

\subsubsection{Futures}

Our futures dataset is obtained from the Pinnacle database, which provides continuously linked contracts (CLC) built from individual futures series across major global exchanges. Pinnacle constructs these continuous series through different adjustment methods, including reverse-adjusted, ratio-adjusted, and non-adjusted splicing, in order to eliminate or preserve contract-to-contract price gaps depending on the research purpose. The database covers a wide range of asset classes—commodities, bonds, equities, and foreign exchange—with data available in both ASCII and Metastock formats at daily frequency.

Each contract includes standardized fields such as open, high, low, settlement price, volume, and open interest. The dataset also offers extended historical coverage, with starting dates ranging from the early 1970s for agricultural futures (e.g., soybeans, corn, wheat) and metals (e.g., gold, silver, copper) to the 1980s and 1990s for financial contracts (e.g., Eurodollars, Treasury bonds, equity indices). Continuous contracts are rolled according to Pinnacle’s rollover schedule, typically occurring in the month prior to delivery, ensuring comparability across maturities.

The futures contracts employed in our empirical analysis are listed in Table~\ref{tab:futures-descriptions}. They span commodities (e.g., CC, KC, ZC, ZW), bonds (e.g., TY, US, UB), currencies (e.g., DX, JN, MP), and equity indices (e.g., ES, NK, YM). This broad coverage enables us to evaluate strategy performance across heterogeneous underlying assets while maintaining consistency in data construction and sampling methodology.

\begin{table}[H]
\centering
\scriptsize 
\setlength{\tabcolsep}{6pt} 
\renewcommand{\arraystretch}{1.05} 
\caption{Description of Futures Contracts Used in the Dataset}
\label{tab:futures-descriptions}
\begin{tabular}{lll}
\toprule
\textbf{Code} & \textbf{Description} & \textbf{Asset Class} \\
\midrule
CC  & Cocoa                     & Commodity \\
DA  & Milk (Class III)          & Commodity \\
GI  & Gold Index                & Commodity \\
JO  & Orange Juice              & Commodity \\
KC  & Coffee                    & Commodity \\
KW  & Hard Red Wheat            & Commodity \\
LB  & Lumber                    & Commodity \\
SB  & Sugar                     & Commodity \\
ZC  & Corn                      & Commodity \\
ZZ  & Gasoline (Unleaded)       & Commodity \\
ZG  & Gold                      & Commodity \\
ZI  & Silver                    & Commodity \\
ZK  & Copper                    & Commodity \\
ZL  & Soybean Oil               & Commodity \\
ZO  & Oats                      & Commodity \\
ZP  & Propane                   & Commodity \\
ZR  & Rough Rice                & Commodity \\
ZU  & Utilities Index           & Commodity \\
ZW  & Wheat                     & Commodity \\
\midrule
ZF  & 5-Year T-Note             & Bond \\
ZN  & 10-Year T-Note            & Bond \\
ZT  & 2-Year T-Note             & Bond \\
CB  & Corporate Bond            & Bond \\
DT  & Eurodollar (Short)        & Bond \\
FB  & Federal Bond Index        & Bond \\
GS  & UK Gilt                   & Bond \\
TU  & 2-Year T-Note             & Bond \\
TY  & 10-Year T-Note            & Bond \\
UB  & Ultra T-Bond              & Bond \\
US  & 30-Year T-Bond            & Bond \\
UZ  & Ultra 10-Year T-Note      & Bond \\
\midrule
AN  & Australian Dollar         & FX \\
CN  & Canadian Dollar           & FX \\
BN  & British Pound             & FX \\
DX  & U.S. Dollar Index         & FX \\
JN  & Japanese Yen              & FX \\
MP  & Mexican Peso              & FX \\
SN  & Swiss Franc               & FX \\
\midrule
FN  & Financial Index           & Equity Index \\
NK  & Nikkei 225                & Equity Index \\
ZA  & South African Index       & Equity Index \\
CA  & Canadian Index            & Equity Index \\
EN  & Euro Stoxx 50             & Equity Index \\
ER  & Russell 2000              & Equity Index \\
ES  & S\&P 500                  & Equity Index \\
LX  & DAX (Germany)             & Equity Index \\
MD  & MidCap 400                & Equity Index \\
XU  & FTSE 100                  & Equity Index \\
XX  & Euro Index                & Equity Index \\
YM  & Dow Jones Industrial Avg  & Equity Index \\
\bottomrule
\end{tabular}
\end{table}

\subsection{Experiment Formulation}\label{sec:experiment_formulation}

To address the main research questions outlined in the introduction, we design two central experiments: one focused on portfolio optimization and the other on time-series momentum.

\subsubsection{Portfolio Optimization}
For the portfolio optimization task, we implement an expanding-window pseudo out-of-sample procedure that recursively solves the optimization problem at each time step. This procedure simulates daily rebalancing, where the optimal portfolio is estimated using information available up to the close of day $t$, applied at the closing price of the same day, and held until the next trading day.

To evaluate the effectiveness of our proposed robust optimization framework, we apply it to the ETF dataset described in the previous section. Performance is assessed in two settings: (i) a long-only allocation, and (ii) a long–short allocation, where portfolio weights may take both positive and negative values.  

We benchmark our methodology against the following strategies:

\begin{itemize}
    \item \textbf{Equal-Weight Portfolio (EW):} A naive diversification rule that allocates an equal share of wealth to each asset in the universe.
    \item \textbf{Mean–Variance Optimization (MVO):} The classical approach of \citet{markowitz-1952}, which maximizes expected return subject to a variance penalty, based on historical estimates of means and covariances.
    \item \textbf{Robust MVO with Uncertainty Sets (RPO):} The distributionally robust model of \citet{ceria-stubbs-2016}, which enhances MVO by introducing ellipsoidal confidence regions around the estimated mean vector.
    \item \textbf{Bootstrap-Based Robust MVO (BUMVO):} Our proposed non-parametric methodology.
\end{itemize}

Robustness and practical effectiveness are assessed through a recursive out-of-sample backtest. At each time $t$, we estimate the parameters required by each method (e.g., the sample mean vector and covariance matrix for MVO) using only data up to $t-1$. These estimates are then used as plug-in inputs to the optimization procedure, and the resulting portfolio weights are applied to compute realized returns from $t$ to $t+1$. This recursive design mimics a realistic investment setting and ensures the absence of look-ahead bias.

Performance is evaluated using standard metrics computed from the backtested returns, conditional on the optimal portfolios obtained at each step:

\begin{itemize}
    \item Expected Return: $\mathbb{E}[R]$
    \item Standard Deviation of Returns: $\mathrm{Std}(R)$
    \item Sharpe Ratio
    \item Sortino Ratio
    \item Average Drawdown (AvgDD)
    \item Maximum Drawdown (MaxDD)
    \item Percentage of Positive Return Periods (\% Positive)
\end{itemize}

\subsubsection{Time-Series Momentum}  
We further assess our framework in a trend-following context using the Time-Series Momentum (TSMOM) strategy of \citet{moskowitz-etal-2012}, applied to both ETF and futures datasets. Each asset is traded based on its own lagged return behavior. At each time $t$, the signal for asset $i$ is defined as the cumulative return over a lookback window $\boldsymbol{\theta}$ (e.g., 3, 6, or 12 months). A positive signal generates a long position, while a negative signal generates a short position. The lookback window $\boldsymbol{\theta}$ is the primary hyperparameter, and our bootstrap-based robust optimization framework is used to select its value by optimizing a functional of a given utility function (e.g., the 25th percentile of the Sharpe ratio across bootstrap samples).

We present results for two utility/loss functions:
\begin{itemize}
    \item Sharpe Ratio
    \item Maximum Drawdown
\end{itemize}

As benchmarks, we consider:
\begin{itemize}
    \item Empirical Risk Minimization (ERM)
    \item Classical Bootstrap on the Asset Level (CB I)
    \item Classical Bootstrap on the Portfolio Level (CB II)
\end{itemize}

Under ERM, $\boldsymbol{\theta}$ is chosen to maximize in-sample utility (e.g., Sharpe ratio) computed directly from the historical data without resampling. Both CB I and CB II select $\boldsymbol{\theta}$ by maximizing utility across bootstrap replications and window lengths, effectively repeating ERM across resamples.

The dataset is divided into training (80\%) and test (20\%) subsamples. The training set is used to perform parameter selection under both our proposed method and the benchmarks, and the selected hyperparameters are then applied to the test set.

In addition to absolute performance, we assess generalization ability through the generalization gap, defined as the difference between the metric evaluated in-sample (train) and out-of-sample (test). This measure captures the extent of overfitting and quantifies the robustness of each method’s hyperparameter choices.

Finally, rather than selecting a single lookback parameter for the entire portfolio, we apply the optimization procedure separately to each asset. This yields one optimal window $\boldsymbol{\theta}$ per asset within the ETF and futures datasets, increasing the effective sample size and allowing for statistical inference across assets. Results are reported in terms of matched utility and loss functions, as well as generalization gaps, with 95\% confidence intervals computed under the Gaussian assumption.

\section{Experiment Results}

This section presents the empirical findings from applying the proposed non-parametric bootstrap-based robust optimization (BUMVO) to the two financial tasks introduced in Section~\ref{sec:benchmarks}: (i) portfolio optimization under both long-only and long-and-short allocation constraints, and (ii) hyperparameter selection for a time-series momentum trading strategy. All results are obtained using the ETF dataset described in Section~\ref{sec:benchmarks} and are computed in fully out-of-sample conditions, with re-estimation and rebalancing performed recursively to avoid look-ahead bias.

\subsection{Portfolio Optimization}

Tables~\ref{tab:long_only_perf} and~\ref{tab:long_short_perf} present a comparative evaluation of portfolio optimization methods under long-only (Table 2) and long–short (Table 3) allocation settings, reporting annualized out-of-sample performance across return and risk metrics. The strategies compared are: equal-weight (EW), classical mean–variance optimization (MVO), robust mean–variance optimization with ellipsoidal uncertainty sets (RPO), and the proposed bootstrap-based robust MVO (BUMVO) evaluated at three robustness percentiles (95, 75, and 25). The metrics include expected return (E[R]), volatility (Std(R)), Sharpe and Sortino ratios, and drawdown measures (average drawdown and maximum drawdown). Bold values denote the best performance within each column.

The results in Table~\ref{tab:long_only_perf} clearly favor bootstrap-based robust MVO. BUMVO\_95 achieves the highest expected return (16.88\%), the strongest Sharpe ratio (0.78), the best Sortino ratio (0.91), and the lowest drawdowns (AvgDD –12.35\%, MaxDD –40.20\%). This dominates the classical benchmarks: MVO, while yielding higher returns than EW (12.22\% vs. 4.75\%), suffers from deeper drawdowns (MaxDD –44.31\%) and a weaker Sharpe ratio (0.58 vs. 0.78 for BUMVO\_95). RPO provides modest improvements over MVO in downside protection (MaxDD –47.37\%), but fails to match BUMVO across metrics. Notably, BUMVO\_75 and BUMVO\_25 also outperform both MVO and RPO on risk-adjusted performance, highlighting the consistency of the bootstrap regularization approach. The equal-weight benchmark, though diversified, is dominated across all dimensions, with the lowest return (4.75\%) and the largest drawdown (–54.41\%).

The long–short setting in Table~\ref{tab:long_short_perf} yields a similar ordering with some nuance. BUMVO\_75 provides the best overall balance, with an expected return of 11.81\% and the highest Sharpe (0.58) and Sortino (0.83) ratios. BUMVO\_95 again delivers the strongest downside protection (AvgDD –12.38\%, MaxDD –38.53\%), though at slightly lower expected return than BUMVO\_75. Classical MVO performs poorly, with lower returns (6.06\%) and the weakest risk profile (MaxDD –60.76\%). RPO achieves higher returns than MVO (9.78\%) and improved risk-adjusted metrics (Sharpe 0.47, Sortino 0.70), but still lags behind BUMVO. EW again underperforms with both low returns and large drawdowns.

Three central findings emerge. First, bootstrap-based robust optimization (BUMVO) systematically outperforms both classical MVO and RPO across return, Sharpe, and drawdown measures. BUMVO\_95 dominates in drawdown control, while BUMVO\_75 offers the best balance between return and risk-adjusted performance, particularly in the long–short case. Second, classical mean–variance optimization is unstable out of sample, suffering from both lower risk-adjusted returns and larger drawdowns, consistent with well-documented criticisms of MVO’s sensitivity to estimation error in expected returns and covariances. Third, the fact that all bootstrap-based variants (BUMVO\_25, BUMVO\_75, BUMVO\_95) consistently outperform their classical counterparts suggests that bootstrap-based robustness is both effective and stable across robustness levels, whereas ellipsoidal uncertainty sets (RPO) provide weaker improvements.

\begin{table}[H]
\centering
\begin{tabular}{lrrrrrr}
\toprule
 & \textbf{E[R]} & \textbf{Std(R)} & \textbf{Sharpe} & \textbf{Sortino} & \textbf{AvgDD} & \textbf{MaxDD} \\
\midrule
ew\_lo & 4.75 & 22.33 & 0.21 & 0.25 & -21.21 & -54.41 \\
mvo\_lo & 12.22 & 20.95 & 0.58 & 0.73 & -17.64 & -44.31 \\
rpo\_lo & 6.75 & 20.94 & 0.32 & 0.39 & -17.94 & -47.37 \\
bumvo\_lo\_95 & 16.88 & 21.68 & \textbf{0.78} & \textbf{0.91} & \textbf{-12.35} & \textbf{-40.20} \\
bumvo\_lo\_75 & 12.90 & 21.45 & 0.60 & 0.72 & -17.23 & -47.14 \\
bumvo\_lo\_25 & 13.40 & 21.00 & 0.64 & 0.77 & -17.57 & -50.24 \\
\bottomrule
\end{tabular}
\caption{Long-only portfolio performance comparison across portfolio optimization methods.  
The table reports annualized out-of-sample performance metrics for the long-only allocation setting.  
Columns correspond to: expected return (E[R], in \%), standard deviation of return (Std(R), in \%), Sharpe ratio, Sortino ratio, average drawdown (AvgDD, in \%), and maximum drawdown (MaxDD, in \%).  
Results are shown for: equal-weight (EW), classical mean--variance optimization (MVO), robust mean--variance optimization with ellipsoidal uncertainty sets (RPO), and the proposed bootstrap-based robust MVO (BUMVO) at three robustness percentiles (95, 75, and 25).  
Bold numbers indicate the best performance in each metric.  
}
\label{tab:long_only_perf}
\end{table}

\begin{table}[H]
\centering
\begin{tabular}{lrrrrrr}
\toprule
 & \textbf{E[R]} & \textbf{Std(R)} & \textbf{Sharpe} & \textbf{Sortino} & \textbf{AvgDD} & \textbf{MaxDD} \\
\midrule
ew\_ls & 4.75 & 22.33 & 0.21 & 0.25 & -21.21 & -54.41 \\
mvo\_ls & 6.06 & 20.55 & 0.29 & 0.40 & -24.06 & -60.76 \\
rpo\_ls & 9.78 & 20.87 & 0.47 & 0.70 & -14.03 & -41.47 \\
bumvo\_ls\_95 & 7.78 & 20.39 & 0.38 & 0.51 & \textbf{-12.38} & \textbf{-38.53} \\
bumvo\_ls\_75 & 11.81 & 20.22 & \textbf{0.58} & \textbf{0.83} & -13.67 & -42.25 \\
bumvo\_ls\_25 & 6.47 & 20.50 & 0.32 & 0.45 & -21.59 & -40.48 \\
\bottomrule
\end{tabular}
\caption{Long-and-short portfolio performance comparison across portfolio optimization methods.  
The table reports annualized out-of-sample performance metrics for the long-and-short allocation setting, where positions can be both positive and negative.  
Columns correspond to: expected return (E[R], in \%), standard deviation of return (Std(R), in \%), Sharpe ratio, Sortino ratio, average drawdown (AvgDD, in \%), and maximum drawdown (MaxDD, in \%).  
Results are shown for: equal-weight (EW), classical mean--variance optimization (MVO), robust mean--variance optimization with ellipsoidal uncertainty sets (RPO), and the proposed bootstrap-based robust MVO (BUMVO) at three robustness percentiles (95, 75, and 25).  
Bold numbers indicate the best performance in each metric.  
}
\label{tab:long_short_perf}
\end{table}

\begin{figure}[H]
    \centering
    \includegraphics[width=0.95\textwidth]{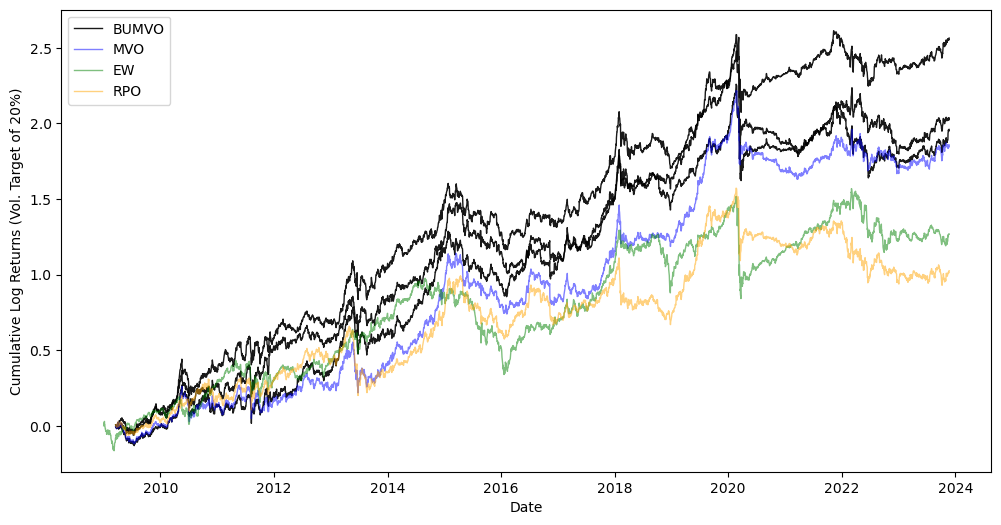}
    \caption{Long-only cumulative portfolio performance comparison across portfolio optimization methods, scaled to a 20\% volatility target. The figure groups strategies by optimization approach (EW, MVO, RPO, and BUMVO) and illustrates their cumulative return trajectories over the evaluation period. The BUMVO variants—particularly at higher percentiles—maintain a consistently steeper growth path, indicating superior compounding and reduced drawdown persistence compared to traditional methods.}
    \label{fig:long_only_group}
\end{figure}

\begin{figure}[H]
    \centering
    \includegraphics[width=0.95\textwidth]{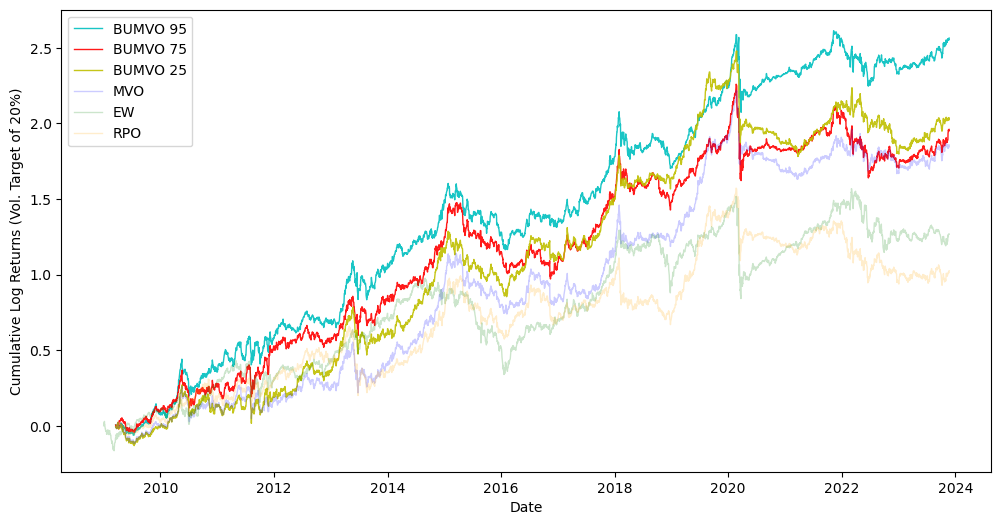}
    \caption{Long-only cumulative portfolio performance comparison across portfolio optimization methods, with a focus on BUMVO percentile configurations (95th, 75th, and 25th percentiles). Lower percentiles represent more conservative optimization, trading off potential return for reduced downside risk. The 95th percentile achieves the strongest overall growth, while the 75th percentile offers a balanced profile between risk and return. All BUMVO configurations outperform standard EW, MVO, and RPO approaches over the out-of-sample horizon.}
    \label{fig:long_only_perc}
\end{figure}

\begin{figure}[H]
    \centering
    \includegraphics[width=0.95\textwidth]{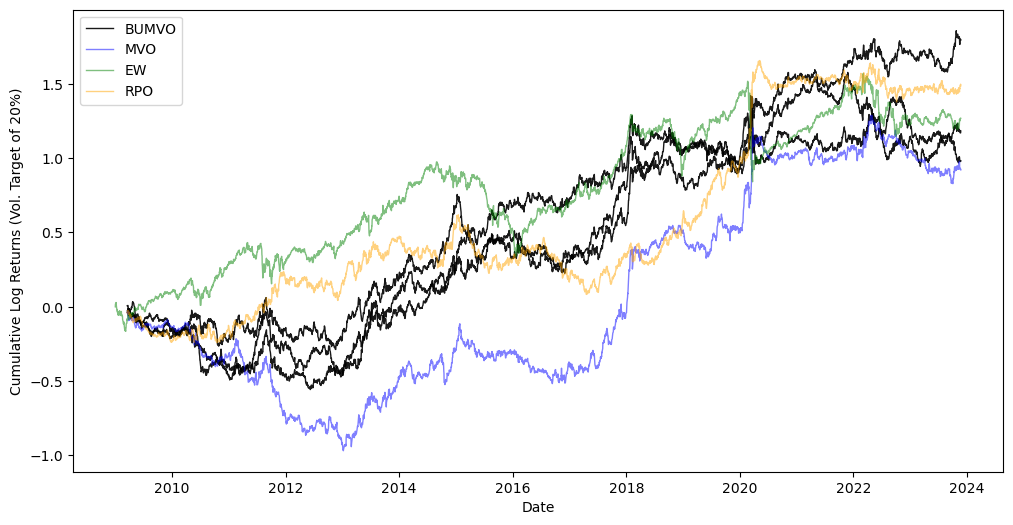}
    \caption{Long-and-short cumulative portfolio performance comparison across portfolio optimization methods, scaled to a 20\% volatility target. The plot groups results by optimization approach, highlighting the performance of EW, MVO, RPO, and BUMVO in a setting that allows both long and short exposures. The BUMVO 75th percentile emerges as the top performer, maintaining a smoother and more resilient growth trajectory than conventional approaches, with the 95th percentile also delivering competitive returns.}
    \label{fig:long_short_group}
\end{figure}

\begin{figure}[H]
    \centering
    \includegraphics[width=0.95\textwidth]{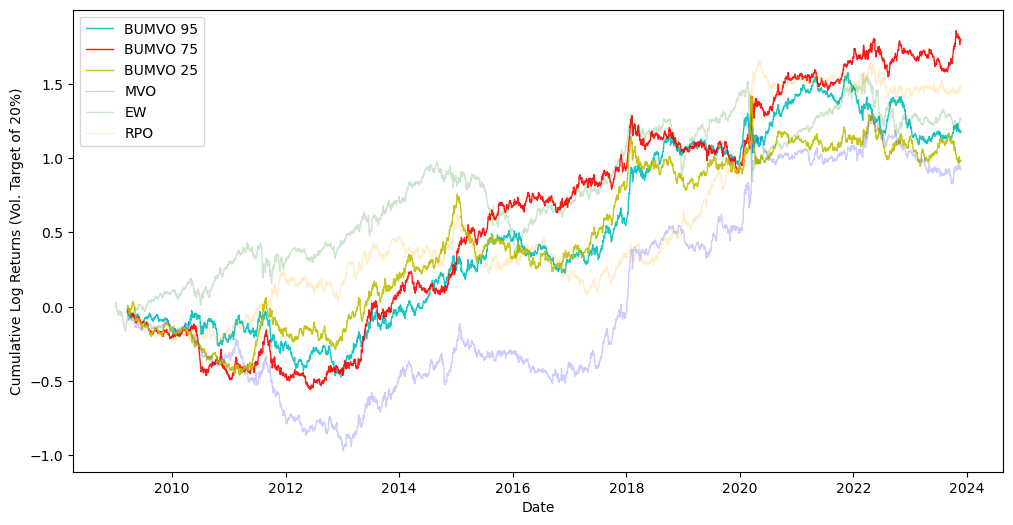}
    \caption{Long-and-short cumulative portfolio performance comparison, focusing on BUMVO percentile configurations. The 75th percentile variant consistently outpaces both the 95th and 25th percentiles, offering a superior balance of return generation and drawdown control. The 95th percentile still delivers robust performance but with slightly higher volatility, while the 25th percentile prioritizes downside protection at the expense of cumulative growth. All BUMVO settings outperform the EW, MVO, and RPO baselines over the evaluation period.}
    \label{fig:long_short_perc}
\end{figure}

\subsection{Time Series Momentum}

\subsubsection{Exchange-Traded Funds}

\begin{figure}[H]
    \centering
    \includegraphics[width=0.85\textwidth]{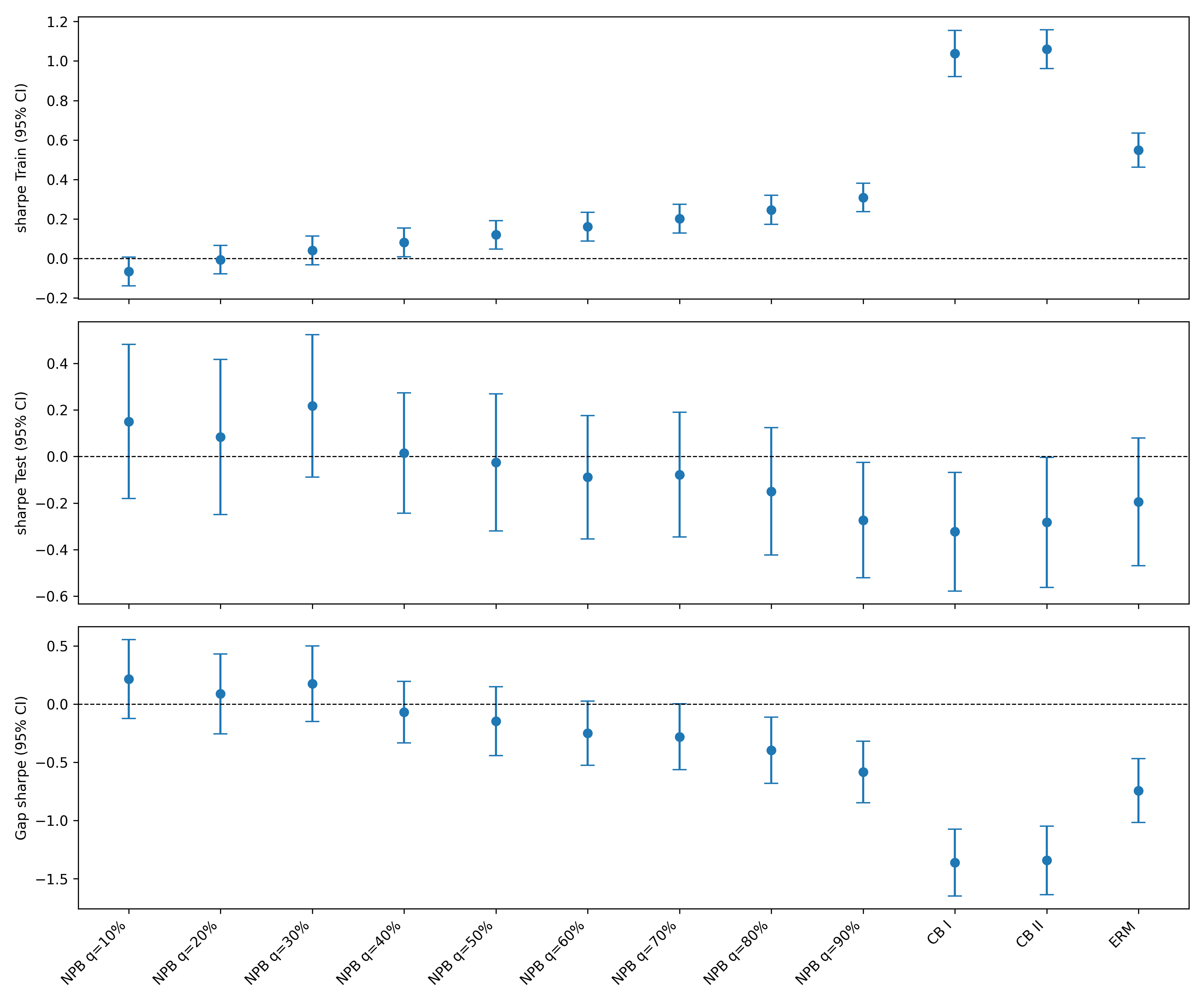}
    \caption{\textbf{Sharpe ratio} with 95\% confidence intervals for the train set (top), test set (middle), 
    and the corresponding generalization gap (bottom) under the \textit{Time-Series Momentum} strategy of \citet{moskowitz-etal-2012}. 
    Parameter values correspond to the lookback window lengths selected via non-parametric robust bootstrap quantiles (10th–90th percentiles) and the ERM\_max benchmark, 
    which maximizes in-sample Sharpe without bootstrap adjustment. 
    Confidence intervals are computed across assets in the \textbf{ETF} dataset under the Gaussian assumption.}
    \label{fig:etfs-moskowitz-sharpe-ci}
\end{figure}

\begin{figure}[H]
    \centering
    \includegraphics[width=0.85\textwidth]{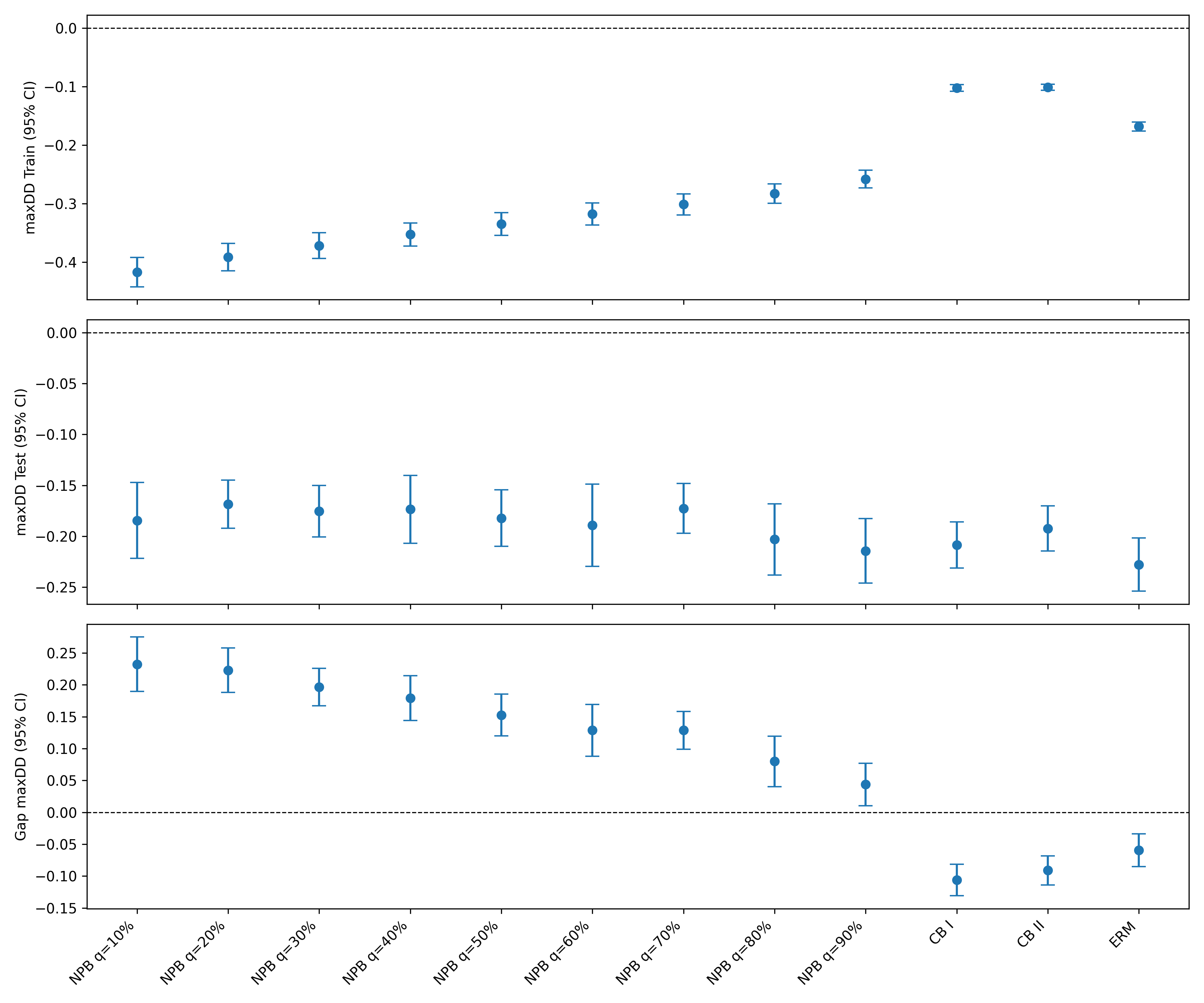}
    \caption{\textbf{Maximum Drawdown} with 95\% confidence intervals for the train set (top), test set (middle), 
    and the corresponding generalization gap (bottom) under the \textit{Time-Series Momentum} strategy of \citet{moskowitz-etal-2012}. 
    Parameter values correspond to the lookback window lengths selected via non-paramertic robust bootstrap quantiles (10th–90th percentiles) and the ERM\_max benchmark, 
    which minimizes in-sample Maximum Drawdown without bootstrap adjustment. 
    Confidence intervals are computed across assets in the \textbf{ETF} dataset under the Gaussian assumption.}
    \label{fig:etfs-moskowitz-maxdd-ci}
\end{figure}

Figures~\ref{fig:etfs-moskowitz-sharpe-ci}–\ref{fig:etfs-moskowitz-maxdd-ci} report out-of-sample evaluation metrics for the Time-Series Momentum (TSMOM) strategy of \citet{moskowitz-etal-2012}, applied to ETFs under varying lookback window lengths. Parameters are selected using non-parametric bootstrap quantiles (10th–90th percentiles) and compared against the ERM, CB I, and CB II benchmarks. Each figure is organized into three panels: training results (top), test results (middle), and the corresponding generalization gap (bottom), with 95\% confidence intervals computed across ETFs under the Gaussian assumption.  

Sharpe ratios (Figure~\ref{fig:etfs-moskowitz-sharpe-ci}) are modestly positive in-sample across most bootstrap quantiles, clustering around 0.1–0.2, while ERM achieves a substantially higher value near 0.6. However, this advantage collapses out of sample: test-period Sharpe ratios become negative for nearly all parameterizations, with ERM performing among the worst. Generalization gaps confirm this sharp deterioration, with declines ranging from –0.1 to –0.6 and approaching –1.0 for ERM, highlighting the overfitting associated with naive in-sample optimization. Importantly, the mid-range bootstrap selections (30th–70th percentiles) exhibit significantly less disappointment out of sample, as reflected by narrower and less negative generalization gaps relative to both extreme quantiles and ERM.  

The two classical bootstrap benchmarks, CB I and CB II, display intermediate behavior but with clear inferiority to the non-parametric bootstrap (NPB) selections. Although both outperform ERM in mean test performance, their 95\% confidence intervals for the Gap Sharpe metric do not intersect with those of any NPB quantiles for either Sharpe ratio or Maximum Drawdown, implying that the differences are statistically significant at the 5\% level. This lack of intersection indicates that the generalization behavior of CB-based methods is distinct and less stable compared to that induced by our robust bootstrap procedure.  

By contrast, the ERM confidence interval for the Gap Sharpe intersects only with the 80th and 90th NPB percentiles, but not with the remaining quantiles, showing that only the most optimistic bootstrap selections are statistically indistinguishable from ERM’s overfit behavior. Lower and mid-range quantiles (10th–70th) are clearly separated, reinforcing the interpretation that moderate ``preparation for disappointment'' leads to more reliable and stable out-of-sample performance.  

Turning to risk, Figure~\ref{fig:etfs-moskowitz-maxdd-ci} shows that maximum drawdowns during training cluster around –30\% across most bootstrap quantiles, while ERM appears artificially superior with an in-sample drawdown closer to –15\%. This advantage, however, vanishes out of sample. In the test period, drawdowns converge to the –18\% to –22\% range for all strategies, with ERM showing the largest deterioration in risk control. The generalization gap reveals the same pattern: bootstrap-based selections maintain relatively moderate and stable increases, while ERM exhibits the sharpest rise—nearly +20 percentage points. Both CB I and CB II again fall well outside the confidence intervals of the NPB quantiles for the Gap MaxDD, indicating statistically significant differences and further emphasizing the robustness of our resampling approach.  

Overall, the ETF results underscore a key principle of robust optimization: preparing to be disappointed is, paradoxically, the best way to avoid disappointment out of sample. In Maximum Drawdown, the relationship is particularly strong, the more conservative (lower-quantile) the bootstrap selection, the more likely it is to yield a pleasant surprise when exposed to new data. In Sharpe ratio terms, the dynamic is more sbtle: moderate caution (mid-quantile selection) delivers the most stable performance, whereas excessive optimism (ERM) or extreme pessimism (low quantiles) both underperform. In summary, incorporating distributional uncertainty through the proposed non-parametric bootstrap, particularly by tempering expectations, proves to be a pragmatic and statistically robust hedge against overfitting.

\subsubsection{Futures}
\begin{figure}[H]
    \centering
    \includegraphics[width=0.85\textwidth]{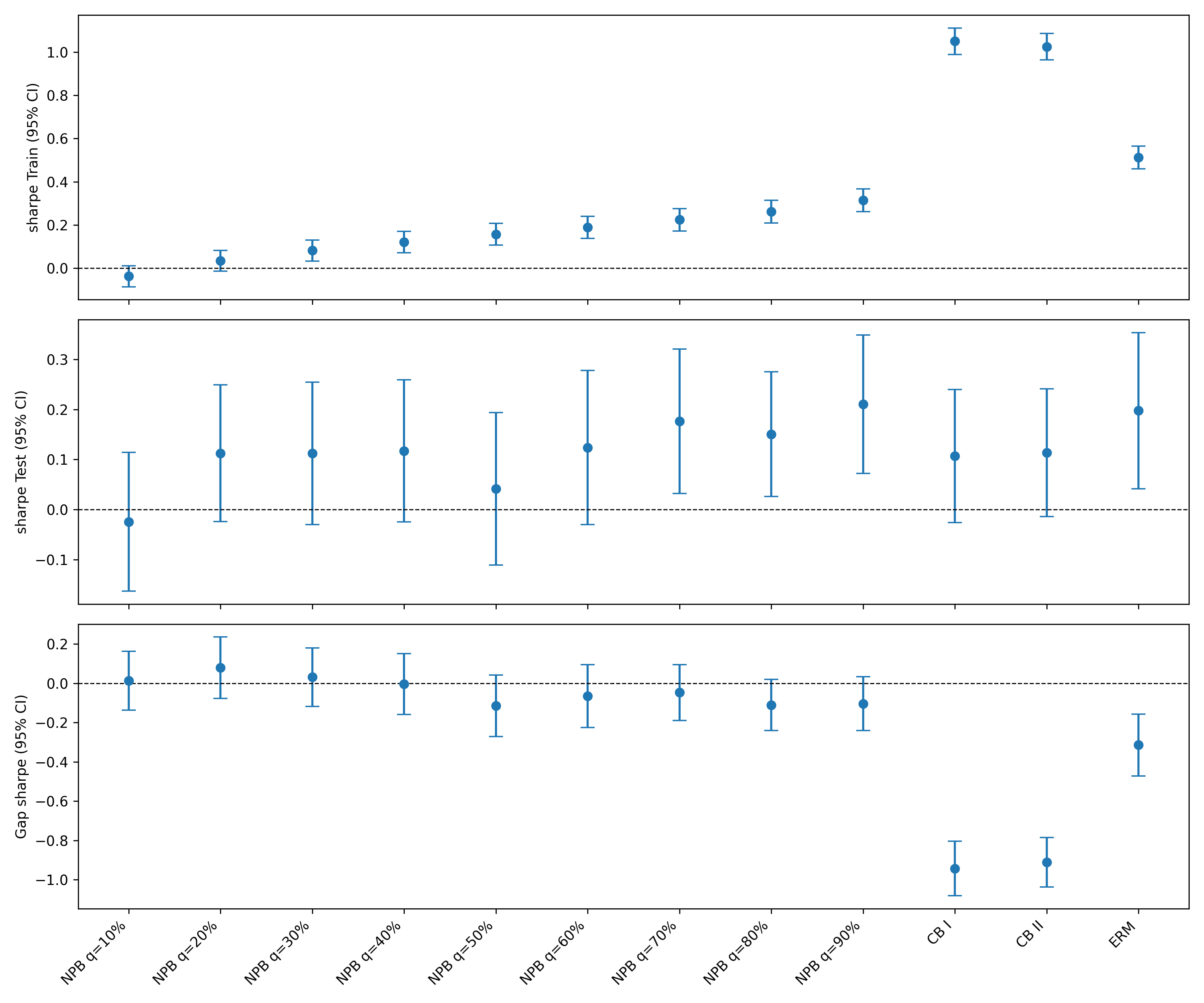}
    \caption{\textbf{Sharpe ratio} with 95\% confidence intervals for the train set (top), test set (middle), 
    and the corresponding generalization gap (bottom) under the \textit{Time-Series Momentum} strategy of \citet{moskowitz-etal-2012}. 
    Parameter values correspond to the lookback window lengths selected via non-parametric robust bootstrap quantiles (10th–90th percentiles) and the ERM\_max benchmark, 
    which maximizes in-sample Sharpe without bootstrap adjustment. 
    Confidence intervals are computed across assets in the $\boldsymbol{\mathrm{Futures}}$ dataset under the Gaussian assumption.}
    \label{fig:futures-moskowitz-sharpe-ci}
\end{figure}

\begin{figure}[H]
    \centering
    \includegraphics[width=0.85\textwidth]{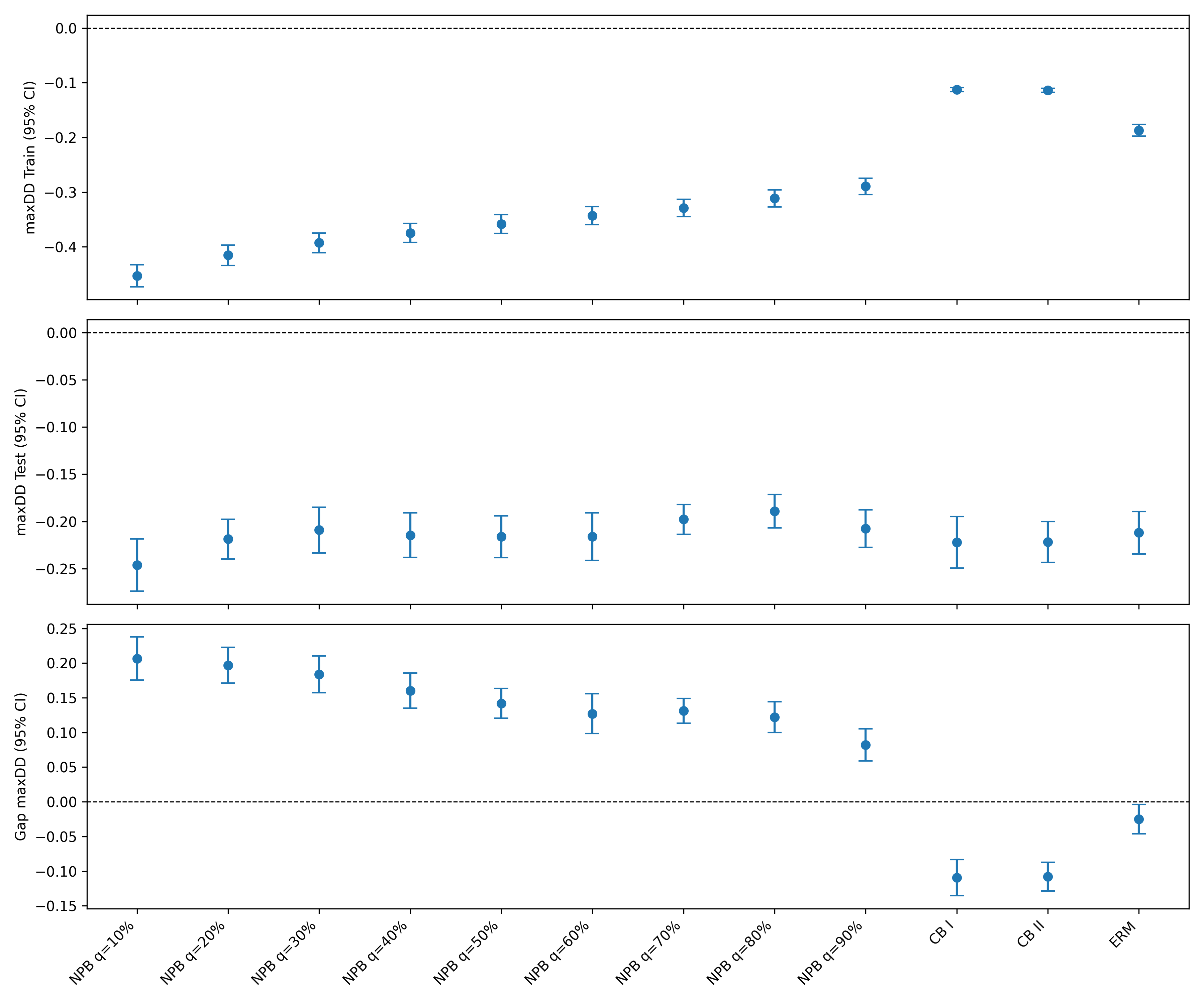}
    \caption{\textbf{Maximum drawdown} with 95\% confidence intervals for the train set (top), test set (middle), 
    and the corresponding generalization gap (bottom) under the \textit{Time-Series Momentum} strategy of \citet{moskowitz-etal-2012}. 
    Parameter values correspond to the lookback window lengths selected via non-parametric robust bootstrap quantiles (10th–90th percentiles) and the ERM\_max benchmark, 
    which maximizes in-sample maximum drawdown without bootstrap adjustment. 
    Confidence intervals are computed across assets in the \textbf{Futures} dataset under the Gaussian assumption.}
    \label{fig:futures-moskowitz-maxdd-ci}
\end{figure}

Figures~\ref{fig:futures-moskowitz-sharpe-ci}–\ref{fig:futures-moskowitz-maxdd-ci} repeat the evaluation for the Futures dataset using the same bootstrap quantiles and benchmark methods (ERM, CB I, and CB II). Results are again presented for the train, test, and generalization gap panels, with 95\% confidence intervals computed across assets under the Gaussian assumption.

Sharpe ratios (Figure~\ref{fig:futures-moskowitz-sharpe-ci}) are higher in-sample than in the ETF experiment, ranging from roughly 0.1 to 0.3 across bootstrap quantiles and nearing 0.6 for ERM. Out-of-sample deterioration is milder, with many strategies maintaining positive Sharpe ratios and confidence intervals that remain above zero. Nevertheless, overfitting remains evident: ERM’s gap approaches –0.5, while the bootstrap-based selections show smaller and more stable declines, generally between –0.1 and –0.4. As in the ETF case, the mid-quantile range (40th–70th percentiles) exhibits the most consistent generalization, with significantly narrower and less negative Gap Sharpe intervals. The 30th, 80th, and 90th percentiles provide modest additional evidence of reduced degradation.  

Confidence interval analysis further strengthens this interpretation. For the Gap Sharpe, the 95\% intervals of CB I and CB II lie entirely outside those of the non-parametric bootstrap (NPB) quantiles, indicating statistically distinct generalization behavior. ERM’s interval overlaps marginally only with the 50th and uppermost quantiles (80th–90th), suggesting that merely the most optimistic bootstrap configurations are statistically indistinguishable from its overfit pattern. These relationships echo the ETF results but with smaller magnitudes, consistent with the greater persistence and smoother trend structure observed in futures markets.

For risk, Figure~\ref{fig:futures-moskowitz-maxdd-ci} shows that training-period maximum drawdowns hover around –30\% for most bootstrap quantiles, while ERM appears artificially favorable at about –20\%. This advantage disappears out of sample: all methods converge to the –18\% to –22\% range in testing. However, the Gap MaxDD highlights clear contrasts. ERM’s gap expands sharply—by nearly +20 percentage points—whereas the bootstrap-based selections exhibit markedly smaller and more stable increases. Confidence intervals for CB I and CB II again lie outside those of the NPB quantiles, supporting statistically significant improvements in out-of-sample risk stability under the robust resampling approach.  

Overall, the Futures results corroborate the patterns observed in ETFs. While ERM and the classical bootstrap methods (CB I and CB II) perform best in-sample, they fail to generalize out of sample, whereas our proposed non-parametric bootstrap (NPB) parameterizations yield smoother and more resilient outcomes. The 40th–70th percentile quantiles once again emerge as the most balanced choices, minimizing both Sharpe ratio degradation and drawdown deterioration. In line with the principle of ``preparing to be disappointed,’’ moderate conservatism consistently leads to more pleasant surprises out of sample, particularly evident in the more persistent futures setting, where robustness is achieved without compromising performance.

\section{Conclusion}

This paper introduces a non-parametric bootstrap framework for robust optimization in finance, designed to mitigate estimation error, parameter uncertainty, and overfitting in portfolio allocation and trading strategy design. Unlike traditional robust optimization methods that rely on restrictive parametric assumptions or ellipsoidal uncertainty sets, our approach leverages data-driven resampling to construct confidence intervals and optimizes utility in a percentile-based manner. This allows the method to flexibly adapt to distributional uncertainty and deliver more stable decision-making under realistic market conditions.

Empirically, the framework delivers consistent gains. In portfolio optimization, bootstrap-based robust MVO (BUMVO) outperforms both classical MVO and ellipsoidal robust optimization, achieving higher risk-adjusted returns and shallower drawdowns in long-only and long–short settings. In time-series momentum, the non-parametric bootstrap markedly improves generalization by tempering the optimism of empirical risk minimization. While ERM and classical bootstrap variants achieve the highest in-sample performance, they suffer from severe out-of-sample degradation. In contrast, mid-percentile selections (around the 40th–70th range) provide smoother, more resilient performance, with smaller generalization gaps in both Sharpe ratio and drawdown metrics. These results demonstrate that incorporating distributional uncertainty through resampling effectively mitigates overfitting and produces more reliable trading strategies.

Several avenues for future research remain. One direction is to extend the robust bootstrap procedure to general hyperparameter tuning problems, particularly in multi-asset trading strategies where parameter instability is pervasive. The choice of quantile threshold within the bootstrap framework also merits deeper theoretical and empirical investigation. Integrating modern sampling techniques, such as generative adversarial networks, may further enrich the modeling of return distributions. Extensions to multi-period portfolio allocation and dynamic trading environments, where risk premia evolve over time, also represent a natural next step. Another promising avenue is to integrate bootstrap-based robustness with Bayesian approaches, combining distributional flexibility with prior information. Finally, theoretical work is needed to clarify the connections between bootstrap-based robustness and worst-case risk minimization frameworks popular in machine learning, and to assess their implications for financial decision-making and hyperparameter optimization.

Taken together, these findings suggest that bootstrap-based robust optimization provides a flexible, data-driven, and practically effective alternative to traditional approaches, with broad potential for advancing both financial econometrics and machine learning applications in trading and portfolio management.

\newpage

\bibliography{_refs}

\end{document}